\documentclass[11pt]{article}
\usepackage{amssymb, latexsym, amsmath, amsthm}
\usepackage{amsfonts, graphics}
\usepackage{url,color}
\usepackage{enumerate}
\def\R{\mathbb R}

\def\A{\mathcal{A}}
\def\p{\left\langle v\right\rangle}

\def\be{\begin{equation}}
\def\ee{\end{equation}}
\def\bea{\begin{eqnarray}}
\def\eea{\end{eqnarray}}
\def\beas{\begin{eqnarray*}}
\def\eeas{\end{eqnarray*}}
\def\supp{\mathrm{supp}\,}

\newcommand{\prfe}{\hspace*{\fill} $\Box$

\smallskip \noindent}

\begin{document}
\sloppy
\newtheorem{theorem}{Theorem}[section]
\newtheorem{definition}[theorem]{Definition}
\newtheorem{proposition}[theorem]{Proposition}
\newtheorem{example}[theorem]{Example}
\newtheorem{cor}[theorem]{Corollary}
\newtheorem{lemma}[theorem]{Lemma}
\theoremstyle{remark}
\newtheorem*{remark}{Remark}

\renewcommand{\theequation}{\arabic{section}.\arabic{equation}}

\title{ On the small redshift limit of steady states of the spherically 
       symmetric Einstein-Vlasov system
       and their stability}

\author{Mahir Had\v{z}i\'{c}\\
        King's College London\\
        Department of Mathematics\\
        Strand\\
        London, WC2R 2LS\\
        email: mahir.hadzic@kcl.ac.uk\\
        \ \\
        Gerhard Rein\\
        Fakult\"at f\"ur Mathematik, Physik und Informatik\\
        Universit\"at Bayreuth\\
        D-95440 Bayreuth, Germany\\
        email: gerhard.rein@uni-bayreuth.de}
        
\maketitle

\begin{abstract}
Families of steady states of the spherically symmetric Einstein-Vlasov system
are constructed, which are parametrized by the central redshift.
It is shown that as the central redshift tends to zero, the states in
such a family are well approximated by a steady state of the Vlasov-Poisson system,
i.e., a Newtonian limit is established where the speed of light is kept
constant as it should be and the limiting behavior is analyzed in terms of a
parameter which is tied to the physical properties of the individual solutions.
This result is then used to investigate the stability properties of the
relativistic steady states with small redshift parameter
in the spirit of recent work by the same authors, i.e.,
the second variation of the ADM mass about such a steady state is shown to
be positive definite on a suitable class of states.  
\end{abstract}

\section{Introduction}

For a relativistic physical theory it is important to understand its
non-relativistic limit. To this end the speed of light $c$ is 
often taken to infinity in the equations of the relativistic theory,
and the corresponding non-relativistic equations pop out.
A more rigorous approach is to prove that solutions of
the relativistic equations converge to solutions of the non-relativistic ones 
in the limit $c\to \infty$. However, both maneuvers are somewhat unsatisfactory,
because in a given set of units the speed of light is a constant.
A more satisfactory approach is to identify families of solutions
of the relativistic theory which are parametrized by a physically
meaningful parameter and to show that these solutions are approximated
well by solutions of the corresponding non-relativistic system when the
parameter tends to some limit.
Following this idea we consider in the present paper 
steady states of the spherically symmetric Einstein-Vlasov system
which we parametrize by the central redshift. We show that they are
approximated well by steady states of the Vlasov-Poisson system
when the central redshift is small. Besides the general issue addressed
above our motivation for this analysis is to deduce stability properties
of the relativistic steady states with small central redshift.
As opposed to the Vlasov-Poisson system the stability properties of steady states
of the Einstein-Vlasov system are mathematically rather poorly understood. 
In \cite{HaRe2013} the present authors study this stability problem 
using the $c\to\infty$ trick and then reinterpreting the result
in terms of the central redshift of the steady states.
The present approach is more direct and at the same time physically 
more meaningful.

Both the Einstein-Vlasov system and the Vlasov-Poisson system
describe large ensembles of particles which
interact only via gravity. 
Galaxies or globular clusters, 
where the stars play the role of the particles, can be modeled as such
ensembles, since collisions among stars are sufficiently rare to 
be neglected. The number density of the 
ensemble on phase space is denoted by $f$, and we assume that all the 
particles in the ensemble have the same rest mass which is normalized to
unity. 
We restrict ourselves to the spherically symmetric, static situation.
For the Einstein-Vlasov system we can therefore write the metric in 
Schwarzschild form
\[
ds^2=- e^{2\mu(r)}dt^2 + e^{2\lambda(r)}dr^2+
r^2(d\theta^2+\sin^2\theta\,d\varphi^2).
\]
Here $t\in\R$ is the time coordinate, $r\in [0, \infty[$ is the
area radius, $\theta\in[0, \pi]$, and $\varphi\in[0, 2\pi]$.
It is useful to introduce the corresponding Cartesian spatial coordinates
\[
x=(r\sin\theta \cos\phi,r\sin\theta \sin\phi,r\cos\theta)\in \R^3.
\]
Asymptotic flatness and
a regular center of the spacetime correspond to the boundary conditions
\be \label{bc}
\lim_{r\to \infty} \lambda(r)= \lim_{r\to \infty} \mu(r) = 0 = \lambda(0).
\ee
The metric coefficients obey the field equations
\begin{equation} \label{eelambda}
e^{-2\lambda} (2 r \lambda' -1) +1 =
8\pi  r^2 \rho,
\end{equation}
\begin{equation} \label{eemu}
e^{-2\lambda} (2 r \mu' +1) -1 =
8\pi  r^2 p,
\end{equation}
The spatial mass-energy density $\rho$ and the radial pressure $p$  are given in terms
of the phase space density $f$ by
\begin{equation} \label{grrhodef}
\rho(r) = \rho(x)
= \int_{\R^3} f(x,v)\sqrt{1+|v|^2} \,dv
\end{equation}
and
\begin{equation} \label{pdef}
p(r) = p(x)
= \int_{\R^3}  f(x,v)\, \left(\frac{x\cdot v}{r}\right)^2\, dv.
\end{equation}
Here $v\in \R^3$ denotes a non-canonical momentum coordinate,
and $f=f(x,v)$ is spherically symmetric,\ i.e., by abuse of notation,
$f(x,v)=f(r,w,L)$ where
\[
r=|x|,\ w=\frac{x\cdot v}{r},\ L=|x\times v|^2;
\]
$|v|$ and $x\cdot v$ denote the Euclidean norm and scalar product respectively.
The fact that the density $f$ on phase space satisfies the Vlasov equation
can equivalently be expressed by demanding that $f$ is constant along
particle trajectories, i.e., along solutions of the characteristic system
\[
\dot x = e^{\mu - \lambda}\frac{v}{\sqrt{1+|v|^2}},\
\dot v = - e^{\mu - \lambda} \sqrt{1+|v|^2} \mu'(r) \frac{x}{r}.
\]
The analogous system in the Newtonian context, 
i.e., the Vlasov-Poisson system, reads
\[
\Delta U = 4 \pi \rho,\ \lim_{|x|\to \infty} U(x) = 0,
\]
\[
\rho (x) = \int_{\R^3} f(x,v)\,dv,
\]
\[
\dot x = v,\
\dot x = - \nabla U (x).
\]
Here $U$ denotes the gravitational potential and
again only the static situation is considered. In 
\cite{RR92b} it was shown that solutions of the
time-dependent Einstein-Vlasov system converge to those
of the Vlasov-Poisson system when the speed of light
is inserted into the former system and is taken to infinity.

In the static situation
the particle energy $E$ defined in \eqref{grparten}
and \eqref{parten} is constant along
characteristics in both cases. Hence the static
Vlasov equation is satisfied if $f$ is taken to be a
function of the particle energy, i.e.,
\be \label{isostate}
f(x,v)=\phi(E).
\ee
This ansatz reduces the corresponding system to the field equation(s)
with source terms which now depend on the metric or the potential.
A fixed ansatz function typically yields a one-parameter family 
of steady states,
and for the case of the Einstein-Vlasov system this can be done in such a way
that the parameter becomes the central redshift $\kappa$ of the resulting galaxy
which is a physically meaningful measure for the strength of relativistic effects
in that steady state. The details of this parametrization are
discussed in Section~\ref{S:STEADYSTATES}.
Our first main result then says that as $\kappa \to 0$
in such a one-parameter family the corresponding steady states
are well approximated by the corresponding
Newtonian steady state. This result is derived in Section~\ref{S:LIMIT}.
The result can then be used to deduce stability properties
for steady states of
the Einstein-Vlasov system which have small central redshift.
The key to stability results for the Vlasov-Poisson system
is to show that on a suitable manifold of dynamically accessible states
at a given steady state
the second variation of the total energy or Hamiltonian
is positive definite.
We deduce the corresponding positivity
result for the Einstein-Vlasov system when the central 
redshift $\kappa$ is small. 
This is in accordance with the
time-honored Zel'dovitch conjecture which says that in such a one-parameter
family of steady states stability holds only as long as the states are
not too relativistic. In Section~\ref{S:STABILITY}
we discuss the general framework for the stability analysis
for the Einstein-Vlasov system and put it into the context of
stability results from the astrophysics literature and in particular
of the work of Zel'dovitch. 
The positive definiteness of the second variation of the Hamiltonian,
i.e., of the ADM mass of the system
is proven in Section~\ref{S:COERCIVITY}.

\section{One-parameter families of steady states}\label{S:STEADYSTATES}
\setcounter{equation}{0}
The problem of constructing static solutions with finite mass
and compact support for the Vlasov-Poisson and Einstein-Vlasov system
has been studied in a number of papers, 
cf.~\cite{RaRe,Rein94,RR93,RR00}. In order to arrive at families
of steady states of the Einstein-Vlasov system which are parametrized
by the central redshift we need to briefly review some of the
corresponding arguments; we also refer to \cite{AnRe1} where such
families and their stability properties were studied numerically.
 
For the static Einstein-Vlasov system the particle energy takes the
form
\be \label{grparten}
E=E(x,v)=e^{\mu(x)}\sqrt{1+|v|^2},
\ee
for the Vlasov-Poisson system the analogous quantity reads
\be \label{parten}
E=E(x,v)=\frac{1}{2} |v|^2 + U(x).
\ee
Since the particle energy in conserved along particle trajectories
any function of the particle energy defines a solution of the
corresponding Vlasov equation if the metric or the potential are viewed
as given.
In the relativistic case we make the ansatz that
\be \label{grtrueansatz}
f(x,v) = \phi(E) = \Phi\left(1-\frac{E}{E_0}\right).
\ee
Here $E_0>0$ is a prescribed cut-off energy---notice that
the particle energy (\ref{grparten}) is always positive---, and
$\Phi$ has the following properties:

\noindent
{\bf Assumptions on $\Phi$.}
$\Phi:\R \to [0,\infty[$ is measurable,
$\Phi (\eta) = 0$ for $\eta < 0$, and for $\eta \to 0+$,
\be\label{E:TAYLOR}
\Phi(\eta) = C \eta^k + \mathrm{O}(\eta^{k+\delta}) 
\ee
with $0<k<3/2$, $\delta >0$, and $C>0$.

This ansatz function $\Phi$ is now kept fixed. We notice that only the metric
quantity $\mu$ enters into the definition of the particle energy
$E$ in \eqref{grparten} and the field equations can be reduced to an
equation for $\mu$. It is tempting to prescribe
$\mu(0)$, but since the ansatz \eqref{grtrueansatz} contains the cut-off
energy $E_0$ as another, in principle free parameter and since
$\mu$ must vanish at infinity due to \eqref{bc} this approach is not
feasible. Instead we define $y:= \ln E_0 - \mu$ so that $e^\mu = E_0 e^{-y}$.  
For the ansatz \eqref{grtrueansatz}
the spatial mass density and pressure become functions of $y$, i.e.,
\be \label{grrhoyrel}
\rho(r) = g(y(r)), \quad
p(r) = h(y(r)),
\ee
where
\be \label{evgdef}
g(y) := \left\{\begin{array}{ccl}
4 \pi e^{3 y} \int_0^{1-e^{-y}} \Phi(\eta)\, (1-\eta)^2\,
\left(e^{2y}(1-\eta)^2-1\right)^{1/2} d\eta&,&y>0,\\
0 &,& y\leq 0,
\end{array}\right.
\ee
and
\be \label{evhdef}
h(y) := \left\{\begin{array}{ccl}
\frac{4 \pi}{3} e^{y} \int_0^{1-e^{-y}} \Phi(\eta)\,
\left(e^{2y}(1-\eta)^2-1\right)^{3/2} d\eta&,&y>0,\\
0 &,& y\leq 0.
\end{array}\right.
\ee
The functions $g$ and $h$ are continuously differentiable 
on $\R$, cf.\ \cite[Lemma 2.2]{RR00}.
The metric coefficient $\lambda$ can be eliminated from the system,
because the field equation \eqref{eelambda} together with the boundary condition
\eqref{bc} at zero imply that
\begin{equation}\label{lambdadef}
e^{-2\lambda(r)} = 1 - \frac{2  m (r)}{r},
\end{equation}
where the mass function $m$ is defined in terms of $\rho$ by
\be \label{mdef}
m(r) = m(r,y) = 4\pi \int_0^r \sigma^2 \rho(\sigma)\, d\sigma.
\ee
Hence the static Einstein-Vlasov
system is reduced to the equation
\be \label{gryeq}
y'(r)= - \frac{1}{1-2 m(r)/r} \left(\frac{m(r)}{r^2} + 4 \pi r p(r)\right) ,
\ee
where $m$, $\rho$, and $p$ are given in terms of $y$ by \eqref{grrhoyrel}
and \eqref{mdef}.

In \cite{RaRe} it is shown that for every central value
\be\label{yiv}
y(0)=\kappa>0,
\ee
there exists a unique smooth solution $y=y_\kappa$
to \eqref{gryeq} which exists on $[0,\infty[$
and which has a unique zero at some radius $R>0$. In view of
\eqref{grrhoyrel}--\eqref{evhdef} this implies that the induced quantities
$\rho$ and $p$ are supported on the interval $[0,R]$, and a non-trivial steady state
of the Einstein-Vlasov system with compact support and finite mass is obtained.
We observe that
the limit $y(\infty):=\lim_{r\to \infty} y(r) < 0$ exists,
the metric quantity $\mu$ is defined by $e^{\mu(r)}= E_0 e^{-y(r)}$, and in order
that $\mu$ has the correct boundary value at infinity we must define 
$E_0=e^{y(\infty)}$. Since $y(R)=0$ we also see that  $E_0=e^{\mu(R)}$.
We want to relate the parameter $\kappa$ to the redshift factor $z$
of a photon which is emitted at the center $r=0$ and received at the
boundary $R$ of the steady state; this is not the standard definition
of the central redshift where the photon is received at infinity,
but it is a more suitable parameter here: 
\[
z = \frac{e^{\mu(R)}}{e^{\mu(0)}} - 1 = \frac{e^{y(0)}}{e^{y(R)}} - 1 = e^\kappa -1.
\]
Hence $\kappa$ is in one-to-one correspondence with the central redshift factor $z$
with $\kappa\to 0$ iff  $z\to 0$, and although this is not the standard terminology
we refer to $\kappa$ as the central redshift.
For a fixed ansatz function $\Phi$ we therefore obtain 
a family $y_\kappa$ of solutions to \eqref{gryeq} and a corresponding family
\[
(f_{\kappa},\lambda_{\kappa},\mu_{\kappa})_{\kappa\in ]0,\infty[}
\]
of steady galaxies to the Einstein-Vlasov system 
parametrized by the central redshift $\kappa$, 
and each member of this family has finite mass
and compact support.

For small $\kappa$ we want to relate the members of the above family to
a suitable steady state of the Vlasov-Poisson system.
For the latter we make the ansatz
\[ 
f(x,v)= \Phi(E_0 - E)
\]
where $\Phi$ has the same properties as above and $E_0<0$.
We define $y=E_0-U$ which satisfies the equation
\be\label{yeq}
y'(r) = - \frac{m(r)}{r^2}
\ee
where the mass function $m$ is defined in terms of $\rho$ as before and
\[ 
\rho(r) = g_0(y(r))
\]
with
\[ 
g_0(y):= \left\{\begin{array}{ccl}
4 \pi \sqrt{2}\int_0^y \Phi(\eta)\, (y -\eta)^{1/2} d\eta&,&y>0,\\
0 &,& y\leq 0.
\end{array}\right.
\]
For every prescribed value $y(0)$ there exists a smooth solution 
of \eqref{yeq} which has a unique zero and gives a steady state
of the Vlasov-Poisson system with finite mass and compact support, 
cf.~\cite{RaRe}.

\section{The small redshift limit}\label{S:LIMIT}
\setcounter{equation}{0}

In order that the solution $y_\kappa$ of \eqref{gryeq}, \eqref{yiv}
converges to some Newtonian limit as 
$\kappa \to 0$ we must properly rescale it.
We define
\[
a = \frac{k+1/2}{2}, 
\]
where we recall that $k$ gives the leading order power in the 
expansion~\eqref{E:TAYLOR} of the profile $\Phi$.
We introduce a rescaled function $\bar{y}=\bar{y}_\kappa$ 
and a rescaled radial variable $s$ by
\be \label{scaling}
y_\kappa (r) = \kappa \, \bar{y}_\kappa (\kappa^a r) 
= \kappa \, \bar{y}_\kappa (s),\quad
s = \kappa^a r.
\ee
Our goal is to derive an equation for the function $\bar{y}$
which corresponds to the equation~\eqref{gryeq} for $y$.
To do so, we introduce the smooth function $F_{\kappa}$ by the relation
\[
e^{-\kappa x} = 1-\kappa x + \kappa^2 F_{\kappa}(x), \ \  x\geq 0
\]
and define
\[
G_{\kappa}(x,\eta) := \eta^2 - F_{\kappa}(2x).
\]
By~\eqref{grrhoyrel}, \eqref{evgdef}, and a change of variables,
\begin{eqnarray*}
\rho(r)  
&=& 
4\pi e^{4\kappa \bar{y}(s)}
\int_0^{\kappa \bar{y}(s)-\kappa^2 F_{\kappa}(\bar{y}(s))}\Phi(\eta) (1-\eta)^2
\left((1-\eta)^2-e^{-2\kappa \bar{y}(s)}\right)^{1/2} \,d\eta\\
&=& 
4\pi \kappa^{3/2} e^{4\kappa \bar{y}(s)}
\int_0^{\bar{y}(s)-\kappa F_{\kappa}(\bar{y}(s))}\Phi(\kappa \eta) (1-\kappa \eta)^2\times \\
&&\qquad \qquad \qquad \qquad
\left(2 (\bar{y}(s)-\eta) + \kappa G_\kappa(\bar{y}(s),\eta)\right)^{1/2} \,d\eta.
\end{eqnarray*}
The analogous computation can be done for the pressure $p$,
and we find the relations
\[
\rho(r) =
\kappa^{3/2} \tilde{\rho}_\kappa(s),\quad
p(r) =
\kappa^{5/2} \tilde{p}_{\kappa}(s),\quad
m(r) = 
\kappa^{3/2-3a} \tilde{m}_{\kappa}(s), 
\]
where
\begin{eqnarray*}
\tilde{\rho}_{\kappa}(s) 
&:=& 
4\pi e^{4\kappa \bar{y}(s)}
\int_0^{\bar{y}(s)-\kappa F_{\kappa}(\bar{y}(s))}\Phi(\kappa \eta) (1-\kappa \eta)^2\times\\
&&
\qquad\qquad\qquad 
\left(2 (\bar{y}(s)-\eta) + \kappa G_\kappa(\bar{y}(s),\eta)\right)^{1/2} d\eta,\\
\tilde{p}_{\kappa}(s) 
&:=& 
\frac{4\pi}{3} e^{4 \kappa \bar{y}(s)}
\int_0^{\bar{y}(s)-\kappa F_{\kappa}(\bar{y}(s))}\Phi(\kappa \eta)
\left(2 (\bar{y}(s)-\eta) + \kappa G_\kappa(\bar{y}(s),\eta)\right)^{3/2} d\eta,\\
\tilde{m}_{\kappa}(s) 
&:=& 
4\pi \int_0^s \sigma^2 \tilde{\rho}_\kappa(\sigma)\, d\sigma.
\end{eqnarray*}
Keeping in mind~\eqref{scaling} we find that 
\[
\frac{m(r)}{r^2}+4\pi r p(r)
 = \kappa^{3/2-a}\left(\frac{\tilde{m}_{\kappa}(s)}{s^2} 
+ 4\pi s\kappa \tilde{p}_{\kappa}(s)\right).
\]
Moreover, 
\[
1-\frac{2 m(r)}{r} = 1- \kappa^{3/2-2a}\frac{2 \tilde m_{\kappa}(s)}{s},
\]
and
\[
y'(r) = \kappa^{1+a} \bar{y}'(s).
\]
Combining the previous three equations we arrive at the 
equation
\[
\bar{y}'(s) = -\frac{\kappa^{1/2-2a}}{1- \kappa^{3/2-2a}\frac{2\tilde{m}_{\kappa}(s)}{s}}
\left(\frac{\tilde{m}_{\kappa}(s)}{s^2} + 4\pi s\kappa \tilde{p}_{\kappa}(s)\right)
\]
satisfied by the rescaled function $\bar{y}$; the corresponding initial
condition becomes $\bar{y}(0) = 1$. 
Finally we define 
\begin{eqnarray}
\Phi_{\kappa}(\eta) 
&:=& 
\kappa^{1/2-2a}\Phi(\kappa \eta),\\
g_{\kappa}(x) 
&:=& 
4\pi e^{4\kappa x}\int_0^{x -\kappa F_{\kappa}(x)} \!\!\!\!\Phi_{\kappa}(\eta)
(1-\kappa\eta)^2\left(2(x-\eta) + \kappa G_{\kappa}(x,\eta)\right)^{1/2} d\eta,\qquad 
\label{gkappadef} \\
h_{\kappa}(x) 
&:=& 
\frac{4\pi}{3} e^{4 \kappa x}\int_0^{x-\kappa F_{\kappa}(x)} \!\!\!\!\Phi_{\kappa}(\eta)
\left(2(x-\eta)+\kappa G_{\kappa}(x,\eta)\right)^{3/2} d\eta, 
\label{hkappadef}
\end{eqnarray}
for $x > 0$, $g_\kappa(x)=h_\kappa(x)=0$ for $x\leq 0$, and
\[
\bar{\rho}_{\kappa}(s) := g_\kappa(\bar{y}_\kappa(s)),\ \quad
\bar{p}_{\kappa}(s) := h_\kappa(\bar{y}_\kappa(s)),\ \quad
\bar{m}_{\kappa}(s) := 
4\pi \int_0^s \sigma^2  \rho_{\kappa}(\sigma)\, d\sigma.
\]
These rescaled functions are related to the original ones by
\be\label{osrel}
\rho_\kappa(r)=\kappa^{1+2a} \bar{\rho}_\kappa(s),\ 
p_\kappa(r)=\kappa^{2+2a} \bar{p}_\kappa(s),\ 
m_\kappa(r)=\kappa^{1-a}\bar{m}_{\kappa}(s).
\ee
We arrive at the following initial value problem which determines the
rescaled function $\bar{y}=\bar{y}_\kappa$:
\be \label{E:MASTER}
\bar{y}'(s) = -\frac{1}{1- \kappa\frac{2\bar{m}_{\kappa}(s)}{s}}
\left(\frac{\bar{m}_{\kappa}(s)}{s^2} 
+ 4\pi \kappa s \, \bar{p}_{\kappa}(s)\right),\quad
\bar{y}(0) = 1. 
\ee
We now show that as $\kappa\to 0$
the solutions of~\eqref{E:MASTER} converge to the solution of the corresponding
Newtonian equation
\be\label{E:LIMIT}
y'(r)  = -\frac{m_0(r)}{r^2},\
y(0)  = 1.
\ee
where 
\[
m_0(r) : = 4\pi \int_0^r\sigma^2 g_0(y(\sigma))\,d\sigma,
\]
\be \label{g0def}
g_0(x) := 4\pi\sqrt{2}  \int_0^x\Phi_0(\eta)(x-\eta)^{1/2}\,d\eta,\ x> 0,
\ee
$g_0(x)=0$ for $x\leq 0$, and 
\[
\Phi_0 (\eta) := \lim_{\kappa\to0}\Phi_{\kappa}(\eta),
\]
more precisely:
\begin{theorem}\label{kappalimit}
Let $\bar{y}_\kappa$ denote the solution of \eqref{E:MASTER} and  
$y_0$ the solution
of  \eqref{E:LIMIT}. There exist constants $\delta>0$, $C>0$, and $\kappa_0>0$
such that for all $0<\kappa\leq \kappa_0$ and $s\geq 0$,
\[
|\bar{y}_\kappa (s) -y_0 (s)| \leq C \kappa^\delta.
\]
\end{theorem}
\noindent
{\bf Proof.}
First we note that the functions $\bar{y}_\kappa$ for $\kappa \geq 0$
are defined on $[0,\infty[$,
and they are decreasing so that $\bar{y}_\kappa(s) \le 1$ for $s\geq 0$
and $\kappa \geq 0$; here we define $\bar y_0 := y_0$.
There exists $S_0>0$ such that $y_0(S_0) < 0$; $S_0$ is strictly to the right
of the support of the Newtonian steady state corresponding to $y_0$.
We aim to show that for $\kappa>0$ sufficiently small, $\bar{y}_\kappa(S_0)<0$
as well.

First we note that there is a constant $C>0$ such that for
all $\eta \in [0,1]$, $\kappa \in ]0,1]$, and $x \in [0,2]$ the estimates
\[ 
|\Phi_\kappa (\eta)|+ |G_\kappa(x,\eta)|\leq C,\ 0 \leq F_\kappa(x) \leq C
\]
hold. This immediately implies that there exists a constant $C_1 >0$ such that
for all $\kappa \in ]0,1]$ and $s\in [0,\infty[$,
\[ 
0\leq \bar{\rho}_\kappa(s),\ \bar{p}_\kappa(s) \leq C_1.
\]
Hence 
\[
\bar{m}_\kappa(s) = 4 \pi \int_0^s \sigma^2 \bar{\rho}_\kappa(\sigma)\, d\sigma \leq 
\frac{4 \pi}{3} C_1 s^3,\ s\geq 0,
\]
and
\be \label{m_est}
\frac{\bar{m}_\kappa(s)}{s}\leq \frac{4 \pi}{3} C_1 S_0^2,\
\frac{\bar{m}_\kappa(s)}{s^2}\leq \frac{4 \pi}{3} C_1 S_0,\ s\in [0,S_0].
\ee
We define
\[ 
\kappa_0 := \left(\frac{16 \pi}{3} C_1 S_0^2\right)^{-1}.
\]
Then \eqref{m_est} implies that for $s\in [0,S_0]$ and 
$\kappa \in ]0,\kappa_0]$,
\[ 
1- \kappa \frac{2 \bar{m}_\kappa(s)}{s} > \frac{1}{2}
\]
and
\be \label{lambda_est}
\left|\frac{1}{1- \kappa \frac{2 \bar{m}_\kappa(s)}{s}} -1\right|
\leq \frac{16 \pi}{3} C_1 S_0^2 \kappa.
\ee
In what follows $C$ shall denote a positive constant which depends 
on the constants appearing above, which may change from line to line, 
and which never depends on $s\in [0,S_0]$ or $\kappa \in ]0,\kappa_0]$.
The estimates which follow hold for all such
$s$ and $\kappa$. Clearly,
\bea
|\bar{y}_\kappa'(s) - y_0'(s)|
&\leq&
\frac{4 \pi s \kappa \bar{p}_\kappa(s)}{1- \kappa \frac{2 \bar{m}_\kappa(s)}{s}}
+ \left|\frac{1}{1- \kappa \frac{2 \bar{m}_\kappa(s)}{s}} -1\right|
\, \frac{\bar{m}_\kappa(s)}{s^2}\nonumber \\
&&
{}+ \left|\frac{\bar{m}_\kappa(s)}{s^2} - \frac{m_0(s)}{s^2} \right|
\nonumber \\
&\leq&
C \kappa + C \int_0^s |\bar{\rho}_\kappa(\sigma) -\rho_0(\sigma)|\, d\sigma.
\label{grw_1}
\eea
Now we recall that $\bar{\rho}_\kappa (s)=g_\kappa(\bar{y}_\kappa(s))$
where $g_\kappa$ is defined in \eqref{gkappadef} for $\kappa>0$
and in \eqref{g0def} for $\kappa=0$.
Under our assumptions on the ansatz function $\Phi$ the function
$g_0$ is continuously differentiable on $\R$, and its derivative is bounded on
$]-\infty,1]$. Hence
\[
\left|\bar{\rho}_\kappa(s) - \rho_0(s)\right| \leq 
\left|g_\kappa(\bar{y}_\kappa(s)) - g_0(\bar{y}_\kappa(s))\right| 
+ C |\bar{y}_\kappa(s) - y_0(s)|.
\]
To complete a Gronwall estimate for the latter difference it remains
to estimate $g_\kappa - g_0$ uniformly on $[0,1]$.
Now
\beas
&&
\left|g_\kappa(x) - g_0(x)\right| 
\leq
4\pi \sqrt{2} \left|e^{4 \kappa x}-1\right| \int_0^{x-\kappa F_\kappa (x)}\ldots \\
&&\qquad\quad
{} + 
4\pi \sqrt{2} \left|\kappa F_\kappa (x)\right| ||\Phi_0||_{L^\infty([0,1])}\\
&&\qquad\quad
{} +
4\pi \sqrt{2}\int_0^{x-\kappa F_\kappa (x)}
\Bigl|\Phi_\kappa(\eta)(1-\kappa\eta)^2 
\left(x -\eta +\kappa \sqrt{2}^{-1} G(x,\eta)\right)^{1/2} \\
&&
\qquad\qquad\qquad \qquad\qquad\qquad \qquad \qquad- 
\Phi_0(\eta) (x-\eta)^{1/2}\Bigr|\, d\eta\\
&&
\qquad \leq
C \kappa + \max_{0\leq \eta \leq x-\kappa F_\kappa (x)} \left|\ldots\right|.
\eeas
By assumption on $\Phi$,
\[
|\Phi_\kappa (\eta) - \Phi_0(\eta)| \leq C \kappa^\delta, \eta\in [0,1].
\]
Moreover,
\[
|(1-\kappa \eta)^2 - 1| \leq C \kappa.
\]
Finally,
\[
\left|\left(x -\eta +\kappa \sqrt{2}^{-1} G(x,\eta)\right)^{1/2}
- (x-\eta)^{1/2}\right| \leq C \kappa^{1/2} |G(x,\eta)|^{1/2}
\leq C \kappa^\delta
\]
where we assume without loss of generality that $\delta \leq 1/2$.
Combining these estimates implies that
\be \label{rhodiffest}
|\bar{\rho}_\kappa(s) - \rho_0(s)| \leq C \kappa^\delta + 
C |\bar{y}_\kappa (s) - y_0(s)|.
\ee
Inserting this into \eqref{grw_1} and integrating in $s$ 
implies that for all $\kappa>0$ sufficiently small and $s\in[0,S_0]$,
\[
|\bar{y}_\kappa (s) -y_0 (s)| \leq C \kappa^\delta 
+ C \int_0^s |\bar{y}_\kappa (\sigma) -y_0 (\sigma)|\, d\sigma
\]
so that by Gronwall's Lemma,
\[
|\bar{y}_\kappa (s) -y_0 (s)| \leq C \kappa^\delta,
\ \ s \in [0,S_0].
\]
In particular, this implies that $\bar{y}_\kappa(S_0) < 0$
for $\kappa>0$ sufficiently small.

We now show that the above estimate remains correct 
for $s\geq S_0$. For $s\geq S_0$ and 
$\kappa \in [0,\kappa_0]$ it holds that
$\bar{\rho}_\kappa(s) = 0$, and hence $m_\kappa (s) = M_\kappa$ is constant.
Hence
\[
y_0(s) = y_0(S_0) + \frac{M_0}{s} - \frac{M_0}{S_0}
\]
and
\[
\bar{y}_\kappa(s) = \bar{y}_\kappa (S_0) 
+ \frac{1}{2\kappa} \ln\left(1-2 \kappa \frac{M_\kappa}{S_0}\right)
- \frac{1}{2\kappa} \ln\left(1-2 \kappa \frac{M_\kappa}{s}\right).
\]
Since $|\bar{y}_\kappa (S_0) - y_0(S_0)| \leq C \kappa^\delta$,
$|M_\kappa - M_0| \leq C \kappa^\delta$, and
\[
\frac{1}{2\kappa} \ln\left(1-2 \kappa \frac{M_\kappa}{s}\right)
= \frac{M_\kappa}{s} + \mathrm{O}(\kappa),
\]
the assertion follows, and the proof is complete.
\prfe

\begin{remark}
Note that the limiting Newtonian polytrope $\Phi_0(\eta) = C \eta^k$ is 
determined solely by the leading order term in the expansion~\eqref{E:TAYLOR} 
of the relativistic profile $\Phi$. This is interesting as it suggests that for 
small central redshifts the steady states to the Einstein-Vlasov system are 
effectively described by the pure Newtonian polytropes $\Phi_0$. 
\end{remark}

The estimate from Theorem~\ref{kappalimit} implies some further
information on the non-relativistic limit $\kappa\to 0$
which will also be needed in Section~\ref{S:COERCIVITY}.
\begin{cor}\label{moreestimates}
Let $\kappa_0 >0$ and $\delta >0$
be as in Theorem~\ref{kappalimit}. There exist
constants $S_0>0$ and $C>0$ such that for all $\kappa \in ]0,\kappa_0]$,
\be \label{uniformsupport}
\supp \bar{\rho}_\kappa \subset [0,S_0],
\ee
and for all $s\geq 0$,
\be\label{scrhoyprconv}
|\bar{\rho}_\kappa (s) - \rho_0(s)| + |\bar{y}'_\kappa (s) - y'_0(s)| \leq C \kappa^\delta.
\ee
In the original, unscaled variables,
\be\label{unscyconv}
\left|\kappa^{-1} y_\kappa (r) -y_0 (\kappa^a r)\right| \leq C \kappa^\delta,
\ee
\be \label{lambdaconv}
\left|e^{2\lambda_\kappa(r)} -1\right|\leq C \kappa,
\ee
and
\be \label{unscrhomuconv}
\left|\kappa^{-1} \mu_\kappa (r) - U (\kappa^a r)\right|+
|\kappa^{-1-2a}\rho_\kappa (r) - \rho_0(\kappa^a r)| \leq C \kappa^\delta.
\ee
\end{cor}
\noindent
{\bf Proof.}
The first assertion follows if we define $S_0$ as in the proof of 
Theorem~\ref{kappalimit} and observe \eqref{gkappadef} together
with the fact that $\bar{y}_\kappa (s)<0$ for $s>S_0$.
If we insert the estimate from Theorem~\ref{kappalimit} into 
\eqref{rhodiffest} we obtain the first estimate in \eqref{scrhoyprconv}.
Inserting this into \eqref{grw_1} yields the second estimate
in \eqref{scrhoyprconv} where we need to observe the uniform control
on the support of $\bar{\rho}_\kappa$.
The estimate \eqref{unscyconv} follows by \eqref{scaling}.
The estimate for $\lambda_\kappa$ follows from
\eqref{lambda_est} and \eqref{osrel}; notice that since
$\bar{m}_\kappa$ is constant for $s\geq S_0$ the estimate
a posteriori holds for all $s\geq 0$. The second estimate in
\eqref{unscrhomuconv} follows again
by \eqref{scaling}. It remains to show the 
estimate for $\mu_\kappa$.
We recall that $\mu_\kappa = \ln E_{0,\kappa}-y_\kappa$
and $U = E_0- y_0$ where the boundary conditions for $\mu_\kappa$
and $U$ imply that $\ln E_{0,\kappa}=\lim_{r\to\infty}y_\kappa(r)$
and $E_0=\lim_{r\to\infty}= y_0(r)$. Hence by \eqref{unscyconv},
\[
|\kappa^{-1}\ln E_{0,\kappa} - E_0| \leq C \kappa^\delta,
\]
and the proof is complete. \prfe

\section{Stability for the Einstein-Vlasov system---a general discussion}
\label{S:STABILITY}
\setcounter{equation}{0}

The study of the dynamic stability of relativistic steady states representing 
static galaxies was initiated in the astrophysics literature in the 1960's, 
first by \textsc{Ze'ldovitch and Podurets}~\cite{ZePo} and then by various other
authors, see~\cite{IT68,IP1969, IP1969b,KM,ZeNo} and the references there.
In these investigations the objects of study are {\em isotropic} steady states 
of the general form \eqref{isostate} which we considered above
where in addition the profile is a decreasing function of the particle energy.
The stability argument in the above works in short is as follows.
A profile $\phi$ is fixed, and by an ad hoc variation of some physical parameter
such as the the central redshift of the galaxy, a one-parameter family of 
steady states is obtained. Then their dynamic stability against spherically 
symmetric perturbations is investigated by solving numerically the  
time-dependent system. 
The remarkable finding, which stands at variance with the analogous situation 
in the Newtonian case (cf.~\cite{GuRe,LeMeRa3}), is that for small values 
of the central redshift $\kappa$, the steady galaxies appear stable 
against spherically symmetric perturbations, but as the value of 
$\kappa$ is increased, the stability changes to instability at some 
critical value $\kappa_{\text{cr}}$. 
In fact, numerical investigations suggest that this exchange is rather 
violent---there exist small perturbations of steady states with 
$\kappa>\kappa_{\text{cr}}$
which appear to lead to gravitational collapse~\cite{AnRe1,ZePo}.
It is the aim of our analysis to better understand this behavior;
we should at this point emphasize that throughout the present paper
we consider only the case of asymptotically flat spacetimes,
which is appropriate for studying isolated systems like
galaxies or globular clusters.

For the stability analysis of a system like the Einstein-Vlasov
system it is essential to understand its conserved quantities.
The flow of the time-dependent problem, which can for example be found in
\cite{HaRe2013}, preserves the 
ADM mass $\mathcal{H}_{\text{ADM}}$ and the Casimir functionals
which act on spherically symmetric phase space densities $f=f(x,v)$: 
\begin{eqnarray}
\mathcal{H}_{\text{ADM}}(f) \label{hdef}
&=&
\iint\sqrt{1+|v|^2} f\,dv\,dx,\\
\mathcal{C}(f)
&=&
\iint e^{\lambda_f}\chi(f)\,dv\,dx, \label{casidef}
\end{eqnarray}
where $\chi \in C^1(\R)$ with $\chi(0)=0$.
If $\chi = \text{Id}$ the associated Casimir is called the particle number:
\be\label{E:PARTICLENUMBER}
\mathcal{N} (f)= \iint e^{\lambda_f}f\,dv\,dx.
\ee
In the above and for what follows it is
important to note that
given a spherically symmetric state $f\in C^1_c (\R^6)$
the metric quantity $\lambda = \lambda_f$ is uniquely
determined by \eqref{lambdadef} and \eqref{mdef}
where the density $\rho$ is defined in terms of $f$ by \eqref{grrhodef};
we occasionally write $\lambda_f$, $\rho_f$, $m_f$ to emphasize
that these quantities are determined by $f$.  
In order to define $\lambda$ by
(\ref{lambdadef}) on $[0,\infty[$
we must require that
\[
 \frac{2 m_f (r)}{r} < 1,\ r\geq 0;
\] 
a non-negative, spherically symmetric state $f\in C^1_c (\R^6)$ 
with this property is called {\em admissible}.

In~\cite{IP1969,IP1969b} the author computes explicitly the second variation 
of the ADM-mass about a given isotropic steady state, inspired by the formalism 
developed by Lynden-Bell for the Vlasov-Poisson system. This leads to a linear 
stability criterion which is then numerically investigated.

Following \cite{AnRe1,KM, ZeNo,ZePo} a related stability conjecture 
can be formulated. The binding energy 
\[
E_b = \frac{\mathcal{H}_{\text{ADM}} - \mathcal{N}}{\mathcal{N}}
\] 
of the steady state as a function of the central redshift $\kappa$ 
has its first maximum at the critical value $\kappa_{\text{cr}}$,
i.e., $\kappa_{\text{cr}}$ is the smallest positive value $\kappa$ where
\[
\frac{d}{d\kappa}E_b(\kappa) = 0.
\]
This condition can physically be interpreted as saying
that once the difference between the total energy and 
the total particle number reaches its first maximum, 
it is advantageous from the energetic point of  view 
to divide up the steady state into $\mathcal{N}$ particles of mass $m=1$ 
thus creating an instability.

From the mathematics point of view
the nonlinear stability properties of relativistic steady states are so far
not understood. Numerical investigations of this question were reported on
in \cite{AnRe1}, and in \cite{Wo} the author employed variational
methods for constructing steady states as minimizers of
certain energy-Casimir type functionals. In~\cite{HaRe2013}, 
the present authors rigorously proved the positive definiteness of the 
second variation of the ADM mass for small parameter values along a  
one-parameter family, parametrized by the speed of light, 
in which the functional dependence~\eqref{isostate} 
is varied by a suitable rescaling. The result is consistent with the 
above discussion in the sense that a posteriori the size of the central 
redshift can be related to the actual parameter, 
but it is not completely in line with the above discussion, where the 
profile $\phi$ (and the speed of light) is fixed and the central redshift 
is varied.

Above we have shown how to parametrize steady states of the Einstein-Vlasov 
system by the central redshift $\kappa$ in such a way that as $\kappa\to 0$
the relativistic steady states are well approximated by a Newtonian one.
This yields uniform estimates, which are used to show the coercivity of the 
second variation 
$D^2\mathcal{H}_{\text{ADM}}(f_\kappa)$ of the ADM mass
on certain dynamically accessible states when $\kappa$ is sufficiently small,
cf.~the next section.
On one hand, this provides a rigorous framework for the mathematical analysis 
of the results in the physics literature explained above, and on the other 
hand, it rigorously confirms the stability findings of 
{\sc Ze'ldovitch} et al.\ at the level of a
linear analysis. However, the coercivity result for the second variation of the 
ADM mass is a {\em nonlinear} result, and it relies crucially on the full 
structure of the Einstein field equations.

For the sake of comparison we recall some facts 
concerning the stability analysis for the Vlasov-Poisson system,
cf.\ \cite{GuRe, LeMeRa3} and the references in the review articles 
\cite{Mou,Rein07}.
For an isotropic steady state $f$
the basic stability condition is that
the dependence on the particle energy is strictly decreasing on the support 
of $f$. If this condition holds all the members of the resulting one-parameter
family of steady states, parametrized by the potential energy difference
between the center and the boundary of the state, are non-linearly stable
against general perturbations. The total energy as well as its
second variation at a given steady state are a-priori indefinite. 
But using the existence of Casimir invariants the dynamics of the system
can be restricted to a leaf $\mathcal{S}_{f}$
of perturbations $g:\R^6\to\R$ which have the same level sets as 
the steady state $f$. On this leaf it is 
possible to establish the positive definiteness of the
second variation of the Hamiltonian about $f$; we refer to \cite{KS}
where this idea appears in the astrophysics literature.
Due to the energy sub-critical nature of the equations, 
such a coercivity estimate 
can be used to prove nonlinear stability.

Given the fact that---as shown in the next section---a corresponding 
coercivity result
for the second variation of the Hamiltonian can also 
be established for the relativistic case
two open problems present themselves. The first one is the question of fully 
nonlinear stability and long-time behavior of the spacetimes generated by 
small perturbations of $f_\kappa$ when $\kappa$ is small. Unlike the 
Vlasov-Poisson system, the Einstein-Vlasov system is energy super-critical, 
and any nonlinear stability analysis will  have to use the structure of the 
system more directly and not just through its conserved quantities. 
The second problem is the dynamic instability 
character of steady states with large $\kappa$ and the study of the ensuing 
gravitational collapse.
The global existence result for the spherically symmetric
Einstein-Vlasov system with small initial data
\cite{RR92a} can be considered as a stability result for the vacuum
solution, but the techniques required for the stability
analysis of non-trivial steady states are 
different from such small data results. On the other hand,
the existing results on gravitational collapse for the Einstein-Vlasov system
so far cover only initial data which are very far from any steady state,
cf.~\cite{AKR1,AKR2,AR} and see also~\cite{DaRe}.

\section{The coercivity estimate}\label{S:COERCIVITY}
\setcounter{equation}{0}

Let $(f,\lambda,\mu)=(f_\kappa,\lambda_\kappa,\mu_\kappa)$ denote 
a steady state solution of the 
Einstein-Vlasov system which is a member of a family as constructed in
Section~\ref{S:STEADYSTATES}; for the moment the dependence
on the central redshift $\kappa$ plays no role and is suppressed. 
We first need to recall
the concept of linearly dynamically accessible perturbations
from \cite{HaRe2013}.
The transport structure of the Vlasov equation imposes a set of natural 
perturbations $(\delta f,\delta \lambda)$ whose defining property is the 
preservation of all Casimir invariants~\eqref{casidef}:
\[
D\mathcal{C}(f)(\delta f)
=\iint e^{\lambda} \left(\chi'(f)\delta f
+\chi(f)\delta\lambda\right) dv\,dx = 0
\]
for all $\chi \in C^1(\R)$ with $\chi(0)=0$, where
\[
\delta\lambda = 
e^{2 \lambda}\frac{4 \pi}{r} \int_0^r s^2 \delta\rho (s)\,ds
\]
and
\[
\delta\rho(r) = \delta\rho(x) = \int\sqrt{1+|v|^2}  \delta f (x,v)\, dv.
\]
As shown in \cite[Thm.~3.2]{HaRe2013}, such perturbations are 
generated by spherically symmetric functions $h\in C^1(\R^6)$ and take 
the form:
\begin{equation} \label{ldynacdef}
\delta f :=
e^{-\lambda}\{h,f\}
+  e^{\mu}\phi'(E)
\frac{w^2}{\sqrt{1+|v|^2}}\delta\lambda,
\end{equation}
where the variation of $\lambda$ is a non-local functional of $h:$
\begin{equation}\label{ldynaclambda}
\delta\lambda=4\pi r  e^{\mu+\lambda}
\int \phi'(E)\,h(x,v)\,w\,dv.
\end{equation}
Here $\phi$ is defined by~\eqref{grtrueansatz} and $\{\cdot,\cdot\}$
denotes the usual Poisson bracket
\[
\{f,g\}:=\partial_x f \cdot\partial_v g-\partial_v f \cdot\partial_x g 
\]
for two continuously differentiable functions $f$ and $g$ of
$x,v \in \R ^3$. The usual product rule for the Poisson bracket reads
\[
\{f,g h\}= \{f,g\} h + \{f,h\} g. 
\]
States of the form (\ref{ldynacdef}) -~\eqref{ldynaclambda} are called
{\em linearly dynamically accessible from $f$}. In what follows we shall
use the abbreviation
\[
\langle v\rangle = \sqrt{1+|v|^2},
\]
and we recall that $w=x\cdot v/r$.

Next we recall from \cite[Eqn.~(2.30)]{HaRe2013} 
the definition the quadratic form
\begin{eqnarray*} 
\mathcal{A}_{\kappa}(\delta f)
&:=&
D^2\mathcal{H}_{\text{ADM}}(f_\kappa)(\delta f,\delta f) \\
&=&
\frac{1}{2}\iint \frac{e^{\lambda_\kappa}}{|\phi'(E)|}(\delta f)^2\,dv\,dx
- \frac{1}{2}\int_0^\infty e^{\mu_\kappa - \lambda_\kappa}
\left(2 r \mu_\kappa' +1\right)\, (\delta\lambda)^2\,dr.\nonumber
\end{eqnarray*}
associated with the ADM-mass;
it should be noted that the parameter $\gamma=1/c^2$ which was
used in \cite{HaRe2013} is equal to unity here.
We can now formulate our main result.
\begin{theorem}\label{th:coercivity}
There exist constants $C^\ast >0$ and $\kappa^\ast>0$ such that
for any $0<\kappa\leq\kappa^\ast$ and any
spherically symmetric function $h\in C^2(\R^6)$ 
which is odd in the $v$-variable the estimate
\[
\mathcal{A}_{\kappa}(\delta f)\geq 
C^\ast\iint|\phi'(E)|\,\left((rw)^2 
\left|\left\{E,\frac{h}{rw}\right\}\right|^2 
+ \kappa^{1+2a} |h|^2\right) dv\,dx
\]
holds. Here $\delta f$ is the dynamically accessible perturbation 
generated by $h$ according to (\ref{ldynacdef}).
\end{theorem}

\noindent
{\bf Proof.}
We first recall some additional information on the Einstein field equations
where we again for the moment suppress the dependence on $\kappa$.
The field equations
\eqref{eelambda} and \eqref{eemu} suffice to determine $\lambda$ and $\mu$, 
but for what follows it is important to note that they
do not constitute the complete set of field equations.
Indeed, if these equations and the Vlasov equation hold, then also
\begin{equation}\label{ee2ndo}
e^{- 2 \lambda} \left(\mu'' + (\mu' - \lambda')(\mu' + \frac{1}{r})\right)
= 4 \pi  q 
\end{equation}
with the tangential pressure $q$ defined by
\[
q(r)=q(x) 
= 
\int  f(x,v) \left|\frac{x\times v}{r}\right|^2
\frac{dv}{\sqrt{1+|v|^2}}. 
\]
If we add the two field equations \eqref{eelambda} and \eqref{eemu}
it follows that
\be \label{lprplusmupr}
\lambda'+ \mu' = 4\pi r e^{2\lambda} (\rho + p) \geq 0.
\ee
Next we recall some auxiliary results from \cite{HaRe2013}. 
By \cite[Lemma~3.3]{HaRe2013} the identity 
\be \label{(i)}
\int \phi'(E) w^2 dv
=
- e^{-\mu}\left(\rho +p\right)
\ee
holds.
According to \cite[Lemma~4.3]{HaRe2013} for
every  spherically symmetric function
$h\in C^2(\R^6)$
the estimate
\be \label{(ii)}
\left(\int |\phi'(E)| |w h|\,dv\right)^2
\leq
e^{-\mu}
\left(\rho + p\right)\int|\phi'(E)|h^2 dv
\ee
holds; here we used \eqref{lprplusmupr}.
Finally,
\be \label{(iii)}
\{E,rw\}
=
e^{\mu}r{\mu}'\p-e^{\mu}\p+\frac{e^{\mu}}{\p}
\ee
cf.~\cite[Lemma~4.4]{HaRe2013}.
As the last preparation for the proof of Theorem~\ref{th:coercivity}
we note that by Theorem~\ref{kappalimit} and 
Corollary~\ref{moreestimates} there exists a constant $C>0$ such that for all
$\kappa\in]0,\kappa_0]$,
\be \label{moreest}
\|\bar{\rho}_{\kappa}\|_\infty,\ \|\bar{p}_{\kappa}\|_\infty,\  
\|\bar{y}_\kappa'\|_\infty,\ \|\lambda_\kappa\|_\infty,\ \|\mu_\kappa\|_\infty \leq C; 
\ee
from \eqref{grrhodef} and \eqref{pdef} it is clear that
$p$ is bounded by $\rho$. Moreover, for all
$(x,v)\in \supp f_\kappa$ the following holds:
\be \label{suppest}
|x| \leq C \kappa^{-a},\ |v|^2 \leq C \kappa
\ee
and
\be \label{yprimesign}
-\frac{\bar{y}'_\kappa(s)}{s} \geq c >0,
\ee
provided $\kappa>0$ is sufficiently small.
As to \eqref{suppest} we recall from \eqref{uniformsupport} 
that in the rescaled variables the spatial support of the steady state
is contained in the interval $[0,S_0]$ with $S_0>0$ independent
of $\kappa$. Together with the scaling \eqref{scaling} this
proves the assertion on the 
spatial support. The properties of the ansatz function $\Phi$ imply that
\[
E/E_0 = e^{\mu_\kappa(r)}\sqrt{1+|v|^2}/E_0 = e^{-y_\kappa(r)}\sqrt{1+|v|^2} \leq 1
\]
on $\supp f_\kappa$ so that by the monotonicity of $y_\kappa$,
\[
\sqrt{1+|v|^2} \leq  e^{y_\kappa(r)} \leq e^\kappa
\]
which yields the bound on $|v|^2$. As to the lower bound on $-\bar{y}'_\kappa(s)/s$
we note that for $s\in ]0,S_0]$,
\[
-\frac{\bar{y}'_\kappa(s)}{s} \geq \frac{\bar{m}_\kappa(s)}{s^3}
\geq  \frac{4\pi}{s^3} \int_0^s\sigma^2 \rho_0(\sigma)\, d\sigma 
- \frac{4\pi}{3}\kappa^\delta.
\]
Given the fact that $\rho_0$ is a continuous, non-negative function
with $\rho_0(0)>0$ the first term is bounded from below by
a positive constant which is independent of $\kappa$,
and \eqref{yprimesign} follows.

We now turn to the proof of the theorem.
For simplicity of notation we drop the index $\kappa$ 
in the notation for steady states as long as this dependence does 
not become essential.
We define
\[
\eta:=\frac{1}{rw}h
\]
which is well defined for $w=0$ 
since $h$ is odd in $v$. By the product rule for the Poisson bracket $\{\cdot,\cdot\}$,
\begin{equation}\label{decomp1}
\{E,h\}=rw\{E,\eta\}+\eta\{E,rw\}.
\end{equation}
Just like in~\cite{HaRe2013} we notice that for a dynamically accessible 
perturbation $\delta f$ defined by~\eqref{ldynacdef} and~\eqref{ldynaclambda} 
we have the decomposition
\begin{equation}\label{a1a2}
2 \A(\delta f)=
\A_1(\delta f)+\A_2(\delta f),
\end{equation}
where 
\begin{eqnarray*}
\A_1(\delta f)
&:=&
\iint e^{-\lambda}|\phi'(E)||\{E,h\}|^2 dv\,dx
-\int_0^{\infty}e^{\mu-\lambda}
(2r{\mu}'+1)(\delta\lambda)^2 dr,\\
&=:&
\A_{11}(\delta f) + \A_{12}(\delta f),\\
\A_2(\delta f)
&:=&
-2 \iint |\phi'(E)|\{E,h\} 
\delta\lambda e^{\mu}\frac{w^2}{\p}\,dv\,dx\\
&&
{}+ \iint|\phi'(E)|e^{2\mu+\lambda}\frac{w^4}{\p^2}(\delta\lambda)^2
dv\,dx \\
&=:&
\A_{21}(\delta f) + \A_{22}(\delta f).
\nonumber
\end{eqnarray*}
It will turn out that $\A_1$ yields the desired
lower bound while $\A_2$ is of higher order in $\kappa$ and can
be controlled by the positive contribution from $\A_1$.

\noindent
{\em Step 1---Estimate on $\mathcal{A}_1.$}
Proceeding just like in the proof of \cite[Thm.~4.2]{HaRe2013}, 
we obtain a lower bound on $\A_1(\delta f):$
\begin{eqnarray*}
\A_1(\delta f)
&\geq& 
\iint |\phi'(E)| e^{-\lambda}(rw)^2|\{E,h\}|^2 dv\,dx\\
&&
{} + \iint|\phi'(E)|e^{2\mu-\lambda} h^2
\left[\mu'' - 3\mu' \lambda' -2 (\mu')^2 +
\frac{2\mu'-\lambda'}{r\langle v\rangle^2}\right]\,dv\,dx;
\end{eqnarray*}
cf.\ \cite[Eqns. (4.9), (4.12)]{HaRe2013}. Using \eqref{ee2ndo} and the fact 
that $q\geq 0$,
\[ 
[\ldots ] \geq -3 (\mu')^2 -2 \mu' \lambda' +\frac{\lambda'-\mu'}{r} 
+ \frac{2\mu'-\lambda'}{r\langle v\rangle^2}.
\]
By \ref{lprplusmupr}, $\lambda' \geq -\mu'$, and hence
\be\label{E:DOTS}
[\dots] \ge -2{\mu}'({\mu}'+{\lambda}')-({\mu}')^2+
\frac{{\mu}'}{r}\left(\frac{3}{\p^2}-2\right)
\ee
We now switch to the rescaled quantities so that
\be\label{E:X}
\mu_\kappa'(r) = - \kappa^{1+a}\bar{y}'(s)
\ee
and by \eqref{lprplusmupr} and \eqref{osrel},
\[
\lambda'(r)+\mu'(r) = 4\pi e^{2\lambda} r(\rho(r)+p(r)) 
= 4\pi \kappa^{1+a}s e^{2\lambda} \left(\rho_{\kappa}(s)+\kappa p_{\kappa}(s)\right).
\]
Going back to~\eqref{E:DOTS} we choose $\kappa$ sufficiently small
so that by \eqref{suppest}, $|v|^2 < 1/4$ and hence
\[
[\ldots] \geq -\kappa^{1+2 a} \frac{\bar{y}'(s)}{s}
\left[\frac{2}{5} - 8\pi \kappa e^{2\lambda}s^2 \left(\bar{\rho}(s) + \kappa \bar{p}\right)
-\kappa s \bar{y}'(s)\right].
\]
The uniform bounds in \eqref{moreest} together with \eqref{yprimesign} imply that for
$\kappa \in ]0,\kappa_0]$ the estimate
\[
[\ldots] \geq C \kappa^{1+2 a}
\]
holds on $\supp f$ with some positive constant $C$, provided $\kappa_0$
is sufficiently small.
We conclude that 
\begin{eqnarray}
 \A_1(\delta f)
&\geq&  
\iint |\phi'(E)| e^{-\lambda}(rw)^2|\{E,\eta\}|^2 dv\,dx \nonumber \\ 
&& 
{}+ C\kappa^{1+2a} \iint|\phi'(E)|e^{2\mu-\lambda}h^2\,dv\,dx. \label{E:CRUCIAL}
\end{eqnarray}

\noindent
{\em Step 2---Estimate on $\mathcal{A}_2.$}
Using the decomposition~(\ref{decomp1}) and keeping in mind 
that $\eta=h/rw$ and the formula (\ref{ldynaclambda}) for $\delta\lambda$
we can rewrite $\A_{21}$ as follows:
\begin{eqnarray*}
\A_{21}
&=&
-8\pi\iint|\phi'(E)|e^{2\mu+\lambda}
\frac{rw^2}{\p}\{E,h\}\left(\int \phi'(E)h\tilde w\,d\tilde v\right) dv\,dx\\
&=&
-8\pi\iint|\phi'(E)|e^{2\mu+\lambda}\frac{r^2w^3}{\p}
\{E,\eta\}\left(\int \phi'(E)h\tilde w\,d\tilde v\right) dv\,dx\\
&&
{}-8\pi\iint|\phi'(E)|e^{2\mu+\lambda}
\frac{w}{\p}\{E,rw\} h \left(\int \phi'(E)h\tilde w\,d\tilde v\right) dv\,dx\\
&=:&
X+Y.
\end{eqnarray*}
Since $\rho(r)+p(r) =\kappa^{1+a} \left(\rho_{\kappa}(s)+\kappa p_{\kappa}(s)\right)$,
the uniform bounds in \eqref{moreest} together with \eqref{(i)}
imply that
\begin{equation}\label{E:helpful2}
\sup_{x\in\R^3} \int|\phi'(E)| w^2\,dv
\leq C \kappa^{1+2a}.
\end{equation}
Using the analogous argument on 
\eqref{(ii)} implies that
\begin{equation}\label{helpful}
\left(\int|\phi'(E)| |wh|\,dv\right)^2
\leq C \kappa^{1+2a} \int|\phi'(E)|h^2\,dv.
\end{equation}
The bounds in \eqref{moreest} and \eqref{suppest},
the Cauchy-Schwarz inequality, and the estimates
(\ref{E:helpful2}), (\ref{helpful})  imply that 
\begin{eqnarray*}
|X|
&\leq& 
C \kappa^{-a} \int\left|\int |\phi'(E)|^{1/2}w 
|\phi'(E)|^{1/2}r w \{E,\eta\}\,dv\right|
\left|\int |\phi'(E)| h w \,dv\right|\,dx \\
&\leq&
C \kappa^{1+a}
\left(\iint |\phi'(E)| |rw\{E,\eta\}|^2 dv\,dx\right)^{1/2}
\left(\iint|\phi'(E)| h^2 dv\,dx\right)^{1/2};
\end{eqnarray*}
in the first step we estimate one factor of $r$
by $C \kappa^{-a}$. Hence by~\eqref{E:CRUCIAL},
\[
|X| \leq C \kappa^{1/2} \A_1.
\]
In order to estimate the term $Y$ we rewrite
the identity \eqref{(iii)}:
\[
\{E,rw\}=e^{\mu}r{\mu}'\p+e^{\mu}\left(-\p+\frac{1}{\p}\right)
=e^{\mu}\left(r{\mu}'-\frac{v^2}{\p}\right).
\]
Recalling~\eqref{E:X} we obtain
\[
|\{E,rw\}|=e^{\mu}\left|- \kappa s \bar{y}'(s)-\frac{v^2}{\p}\right| \le C \kappa, \  \
(x,v)\in\supp f,
\]
where we used \eqref{suppest}.
Using this together with the estimates \eqref{moreest}
we proceed as above to find that
\begin{eqnarray*}
|Y|
&\leq&
C\kappa^{3+2a}
\left(\iint |\phi'(E)| h^2\,dv\,dx\right)^{1/2}
\left(\iint|\phi'(E)| h^2 \,dv\,dx\right)^{1/2} \\
&\leq&
C \kappa^2 \A_1.
\end{eqnarray*}
From the above estimates for $|X|$ and $|Y|$
it follows that
\[
\A_2\geq -|X|-|Y| \geq -C \kappa^{1/2} \A_1,
\]
and using (\ref{a1a2}) we finally infer that
\[
\A \geq \frac{1}{2}\A_1-C \kappa^{1/2} \A_1 \geq \frac{1}{4}\A_1,
\]
provided the central redshift $\kappa$ is sufficiently 
small. In view of (\ref{E:CRUCIAL}) the proof is complete. 
\prfe

Theorem~\ref{th:coercivity} in particular implies that 
the steady states of the Einstein-Vlasov system with small values of the 
central redshift $\kappa$ are linearly stable 
against  linearly dynamically accessible perturbations. 
This follows from the fact that the quadratic 
form $\mathcal{A}$ is a conserved quantity along the linearized dynamics. 
The proof is completely analogous to the one 
of \cite[Thm.~6.2]{HaRe2013} and is left out.

\end{document}